\documentclass[twocolumn]{aastex6}

\slugcomment{Accepted by ApJ, April 2019}
\usepackage{natbib,color,graphicx}
\usepackage{textcomp}
\usepackage{float}
\usepackage{soul}
\usepackage{enumerate}
\usepackage{amsmath}
\usepackage{lineno}

\shorttitle{lensing of solar radio emission: spectral caustics}
\shortauthors{Koval et al.}

\begin{document}

\title{Direct Observations of Traveling Ionospheric Disturbances \\ as Focusers of Solar Radiation: Spectral Caustics}

\author{Artem Koval\altaffilmark{1}, Yao Chen\altaffilmark{1}, Takuya Tsugawa\altaffilmark{2}, Yuichi Otsuka\altaffilmark{3}, Atsuki Shinbori\altaffilmark{3}, Michi Nishioka\altaffilmark{2}, Anatoliy Brazhenko\altaffilmark{4}, Aleksander Stanislavsky\altaffilmark{5}, Aleksander Konovalenko\altaffilmark{5}, Qing-He Zhang\altaffilmark{1}, Christian Monstein\altaffilmark{6}, Roman Gorgutsa\altaffilmark{7}}

\altaffiltext{1}{Shandong Provincial Key Laboratory of Optical Astronomy and Solar-Terrestrial Environment, and Institute of Space Sciences, \\ Shandong University, Weihai, China; art$\makebox[0.1cm]{\hrulefill}$koval@yahoo.com}
\altaffiltext{2}{National Institute of Information and Communications Technology, Tokyo, Japan}
\altaffiltext{3}{Institute for Space-Earth Environmental Research, Nagoya University, Nagoya, Japan}
\altaffiltext{4}{Gravimetrical Observatory, S. Subbotin Institute of Geophysics, Poltava, Ukraine}
\altaffiltext{5}{Institute of Radio Astronomy, National Academy of Sciences of Ukraine, Kharkiv, Ukraine}
\altaffiltext{6}{Istituto Ricerche Solari Locarno, Locarno, Switzerland}
\altaffiltext{7}{Pushkov Institute of Terrestrial Magnetism, Ionosphere and Radio Wave Propagation Russian Academy of Sciences, Troitsk, Russia}

\begin{abstract}

The solar radiation focusing effect is related to the specific phenomenon of propagation of the Sun-emitted HF and VHF waves through terrestrial ionosphere. This natural effect is observed with ground-based radio instruments running within 10-200 MHz range, as distinctive patterns -- the Spectral Caustics (SCs) -- on the solar dynamic spectra. It has been suggested that SCs are associated with medium-scale traveling ionospheric disturbances (MSTIDs). In this paper, we present the first direct observations of SCs induced by MSTIDs, using solar dynamic spectra with SCs obtained by different European radio telescopes on January 8, 2014 and simultaneous two-dimensional detrended total electron content (dTEC) maps over Europe. Spatial examination of dTEC maps as well as precise timing analysis of the maps and the dynamic spectra have been performed. First, we found several pairs of one-to-one (TID-SC) correspondences. The study provides strong observational evidence supporting the suggestion that MSTIDs are the cause of SCs.

\end{abstract}


\section{Introduction}
\label{Sect:Introduction}

The focusing effect of terrestrial ionosphere on the solar radiation unmasks itself in ground-based observations of the Sun by radio telescopes in meter and decameter wavelengths. Its manifestations are called as Spectral Caustics (SCs). They emerge in the solar dynamic spectra and exhibit unusual morphology and specific time-frequency features. The interpretation of SCs in terms of diffraction and focusing of solar radiation on ionospheric irregularities has been given by \citet{Meyer-Vernet1980, Meyer-Vernet1981}. Keen interest on this subject was demonstrated in following observational works by \citet{Bougeret1981}, \citet{Genova1983}, \citet{Mercier1986a, Mercier1986b}, and \citet{Mercier1989}. In particular, spectral and heliographic methods for further exploration of SCs were used. In later years, there appeared a long lull in the investigation of the focusing effect of solar radio emission.

From a careful review of earlier studies we concluded that the SC topic has not been fully explored yet. Therefore, we have revisited this subject by carrying out a statistical analysis of SCs in the solar dynamic spectra observed by the Nan\c{c}ay Decameter Array \citep[NDA;][]{Lecacheux2000} covering 10-80 MHz band \citep{Koval2017}. Over a 17-year period (1999-2015) we identified SCs in 129 observational days. For the first time, SCs have been classified into several types according to their spectral manifestations. There are inverted V-like, V-like, X-like, fiber-like, and frindge-like types of SCs, with rare exceptions (i.e., unclassified events). Also, we established seasonal and solar cycle dependencies in the occurrence rate of SCs. In particular, most SCs appear during late fall, winter, and early spring \citep[see details in][]{Koval2017}.

It is suggested that ionospheric irregularities accounting for SCs are traveling ionospheric disturbances (TIDs) of medium scale (MSTIDs), however, the origin of SCs may still cause disputes among solar radio observers. TIDs are spatial quasi-periodic electron density structures produced by acoustic gravity waves (AGWs) \citep{Hines1960}. MSTIDs are characterized by spatial and temporal periods of 100-600 km and 0.25-1 h, respectively \citep{Hunsucker1982}. They can propagate as the guided wave mode in a mirror-like layer at heights 250-300 km above the ground \citep{Fedorenko2013}. Because of its nature the density perturbation consists of alternating structural cells with high and low magnitudes of electron concentration. Since a plasma depletion cell holds higher value of refractive index as compared to its adjoining cells, the cell can cause refraction and focusing of incident electromagnetic wave if plasma density gradients are large enough. As a result, the caustic surfaces produced by MSTIDs can appear in space, and transform into SCs on the solar dynamic spectra. The SCs modeling on the basis of geometric optics has been performed in our latest study \citep{Koval2018}.

In the study we have simulated the propagation of a plane electromagnetic wave through the perturbed ionosphere with MSTIDs which was approximated by the electron density model given by \citet{Hooke1968}, using a two-dimensional ray-tracing method. The method can reveal the trajectories of radio rays through the ionosphere irregularities, and thus the formation of caustics in space under different conditions. We could simulate four types of SCs among the five ones from our classification. This covers the majority of the observed SC morphologies. Also, we gave an explanation of the seasonal behavior in the SC emergence. According to our simulations SCs are generated mainly at relatively low solar elevation angles ($<30^\circ$). This corresponds to late fall, winter, and early spring in the Northern Hemisphere, in line with our earlier statistical study.

Despite these latest studies and earlier ones \citep{Meyer-Vernet1980,Meyer-Vernet1981,Koval2017,Koval2018}, a major challenge remains, which is to capture observationally a specific TID perturbation caused a specific SC. In other words, can we identify a one-to-one -- MSTID-SC -- correspondence using available data? So far, direct observations of a TID as the cause of a SC have never been reported. To address this challenge we analyze the simultaneous-observed solar dynamic spectra and detrended total electron content (dTEC) maps, which have been used as an effective tool in TIDs studies \citep{Tsugawa2007, Otsuka2013} since pioneering work of \citet{Saito1998}. An event on 2014 January 8 has been selected. The following section~\ref{Sect:Observations and Methods} describes the instruments, data and methodology, our results are given in section~\ref{Sect:Results}, and the latest section~\ref{Sect:Summary and Discussion} presents the summary and discussion.

\section{Instruments, Observational Data and Methodology}
\label{Sect:Observations and Methods}
\subsection{Solar Radio Instruments}

The solar radio observations on 2014 January 8 were conducted by many ground-based instruments in Europe. We picked out only those records with clear SCs, by the following radio astronomical facilities:
\begin{itemize}
\item NDA (location: Nan\c{c}ay, France; coords: $\phi_0 = 47.38^\circ$, $\lambda_0 = 2.193^\circ$; frequency band: 10-80 MHz; aerial: phased array consisting of 144 elements)) \citep{Lecacheux2000};

\item URAN-2 (location: Stepanovka, Ukraine; coords: $\phi_0 = 49.63^\circ$, $\lambda_0 = 34.825^\circ$; frequency band: 8-33 MHz; aerial: phased array consisting of 512 elements)) \citep{Konovalenko2016};

\item IZMIRAN observatory (location: Troitsk, Russia; coords: $\phi_0 = 55.482^\circ$, $\lambda_0 = 37.31^\circ$; frequency band: 25-270 MHz; aerial: 10-meter dish, 2 separate dipole antennas) \citep{Gorgutsa2001};

\item CALLISTO BIR (location: Birr, Ireland; coords: $\phi_0 = 53.094^\circ$, $\lambda_0 = -7.92^\circ$; frequency band: 10-105 MHz; aerial: bicone antenna);

\item CALLISTO BLENSW (location: Bleien, Switzerland; coords: $\phi_0 = 47.34^\circ$, $\lambda_0 = 8.112^\circ$; frequency band: 8-79.6 MHz; aerial: bicone antenna);

\item CALLISTO DARO (location: Dingden, Germany; coords: $\phi_0 = 51.77^\circ$, $\lambda_0 = 6.623^\circ$; frequency band: 20-80 MHz; aerial: tilted terminated folded dipole antenna);

\item CALLISTO ESSEN (location: Essen, Germany; coords: $\phi_0 = 51.394^\circ$, $\lambda_0 = 6.979^\circ$; frequency band: 30-90 MHz; aerial: bicone antenna);

\item CALLISTO GLASGOW (location: Glasgow, Scotland; coords: $\phi_0 = 55.902^\circ$, $\lambda_0 = -4.307^\circ$; frequency band: 45-80.8 MHz; aerial: log-periodic dipole array antenna).
\end{itemize}

The NDA and the URAN-2 are phased arrays which have automatic solar tracking systems. The angular sizes of the NDA and the URAN-2 antenna patterns are about $6^\circ\times10^\circ$ and $3.5^\circ\times7^\circ$, correspondingly, at 25 MHz. The IZMIRAN's dish antenna and separate dipoles also track the Sun. They are regularly pointed with steering commands from an observer. The beam width of the 10-meter dish is near $20^\circ$ within the 90-270 MHz range, while dipoles, running in 25-50 MHz and 45-90 MHz bands, have beam width about $40^\circ$. The above-listed CALLISTO devices (spectrometer and antenna) have single and non-tracking antennas with beam width reaching about several tens of degrees (\citep{Benz2005}; \url{http://www.e-callisto.org/}).

\subsection{Detrended TEC maps}

\begin{figure}[t]
\begin{center}
\includegraphics[width=0.5\textwidth]{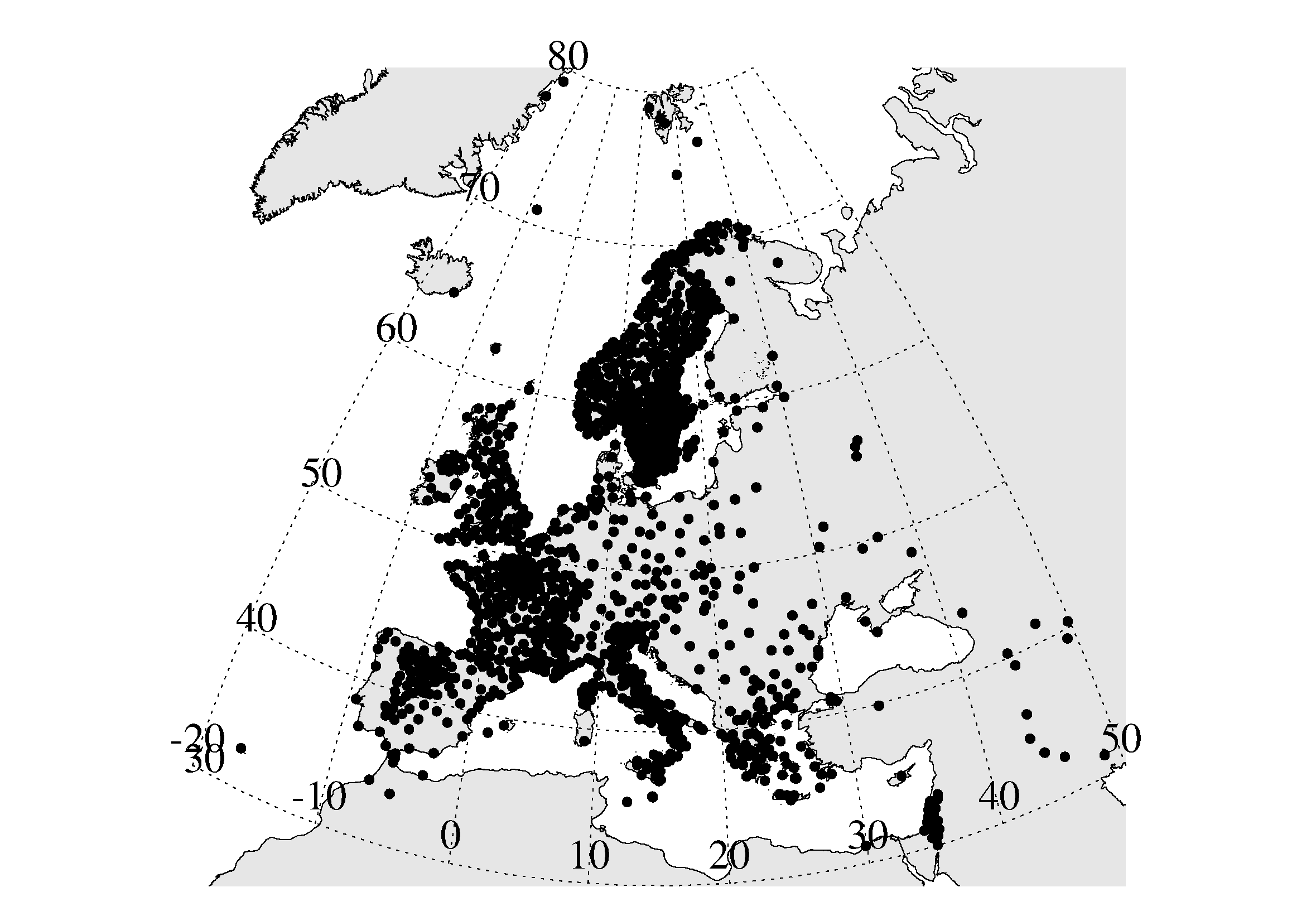}
\caption{Distribution of GPS stations over Europe on January 8, 2014.}
\label{Figure1}
\end{center}
\end{figure}

The TEC data analyzed in this study have been obtained with the GPS receivers in Europe. The data from 1617 permanent GPS receivers on January 8, 2014 have been gathered. Its locations are shown in Figure~\ref{Figure1}. The measurements of the pseudo-ranges and carrier phases are updated at least every 30 s by each dual-frequency GPS receiver working at $f_\textrm{1}$ = 1575.42 MHz and $f_\textrm{2}$ = 1227.60 MHz. Slant TEC, being integrated electron density along the path between receiver and satellite, is obtained from the carrier phase advance. If only the perturbation component of TEC caused by AGWs is required, as is the case of the present work, it is not necessary to determine the absolute TEC value.

In this study, we obtained the perturbation component of TEC by detrending the vertical TEC time series with 1-h running average for each satellite-receiver pair. This method is appropriate for dense nets of GPS stations and has been extensively used \citep{Tsugawa2018}. To compute the vertical TEC, the slant TEC was multiplied by the slant factor. The latter is a ratio of the ionosphere thickness (250~km) for the zenith path to the signal path length between 200 and 450 km of altitude. In the slant-to-vertical TEC conversion routine the TEC data from satellites at zenith angles $\leq55^\circ$ have been used.

The two-dimensional dTEC maps over Europe have a latitude-longitude resolution of $0.15^\circ\times0.15^\circ$. Each pixel represents the average obtained from dTEC values of satellite-receiver pairs whose signal paths cross the pixel at the altitude of 250 km. To get denser filling in each dTEC map, we applied spatially smoothing within $5\times5$/$\cos(\phi)$ pixels in latitude and longitude, where $\phi$ is latitude in degrees \citep[see details in][]{Tsugawa2007}. Consequently, dTEC distributions are mapped to a geographical map at an altitude of 250 km with spatial resolution of $0.75^\circ\times0.75^\circ$/$\cos(\phi)$ ($\thicksim80$~km $ \times$ 80~km). The dTEC values have been obtained in an area from $20^\circ$ W to $50^\circ$ E in longitude and from $30^\circ$ N to $75^\circ$ N in latitude.


\subsection{The TID-SC pair identification methodology}

The goal of our study is to pinpoint a ``parental'' TID that causes a SC. To implement this, we adopt the method which is generally used when building TEC maps. The ionosphere is simplified as a thin layer (shell). The altitude of the shell is generally set equal to the height of the F2 peak (hmF2). The integrated magnitude of electron density within the F2 ionospheric layer is the main contribution to TEC values. Then, the obtained TEC values are given to the so-called ionospheric pierce point (IPP), and mapped on the shell. The IPP represents the point at which the satellite-receiver line-of-sight crosses the ionosphere shell.

Let us substitute a satellite and a receiver for the Sun and a radio telescope, respectively. Therefore, we can localize the place -- the IPP position -- where the Sun-antenna line-of-sight pierces the ionosphere shell. In this way, within the life time of a particular SC, a set of IPP coordinates can be determined and plotted onto the dTEC maps. This allows us to directly observe the instantaneous dTEC structure and the position at which the wave front of incident radio wave penetrates the ionosphere. For ease, we term the TID wavefront as a crest and the region between two successive wavefronts as a valley. The crests and valleys have high and low magnitudes of electron density, respectively. The focusing could only happen when the solar radio wave propagates through a valley where the refractive index is higher than that in the surrounding crests. Therefore, we expect to find that the IPP positions corresponding to any SC being contained within a valley on dTEC map. If so, we would be able to identify the particular TID responsible for the specific SC.

The geographical latitude ($\phi_I$) and longitude ($\lambda_I$) of an IPP can be computed using following formulas given by \citet{Klobuchar1987}:
\begin{eqnarray}
\begin{gathered}
\beta = 90 - \theta - \arcsin\left[\frac{R_E}{R_E+h}\cos\theta\right],\\
\phi_I = \arcsin(\sin\phi_0\cos\beta + \cos\phi_0\sin\beta\cos A),\\
\lambda_I = \lambda_0 + \arcsin\left(\frac{\sin\beta\sin A}{\cos\phi_0}\right),
\label{Eq1}
\end{gathered}
\end{eqnarray}
where $R_E = 6371$ km is the radius of the Earth, and $h$ is the height of the ionosphere surface. These formulas are used to calculate the IPP coordinates for a satellite-receiver pair. In our case, $\phi_0$ and $\lambda_0$ are latitude and longitude of the site of a radio telescope, respectively, while $\theta$ is the solar elevation angle and $A$ is the solar azimuth angle.

It is important to define correctly the height of the ionosphere shell $h$. We set it to be 250 km. The same value was taken for dTEC maps. To assign it, we have crosschecked two ionospheric models, the International Reference Ionosphere Extended to the Plasmasphere (IRI-Plas 2017; \url{http://www.ionolab.org/}; \citet{Sezen2018}) and the International Reference Ionosphere (IRI 2016; \url{https://ccmc.gsfc.nasa.gov/}). We applied the global instantaneous ionospheric maps of hmF2 in area covered by dTEC maps. The hmF2 magnitude over a large area containing the IPP tracks (see below) fluctuates only slightly around the value at the altitude of 250 km during our observations.

\section{Results}
\label{Sect:Results}

\begin{figure*}[h]
\begin{center}
\includegraphics[width=1\textwidth]{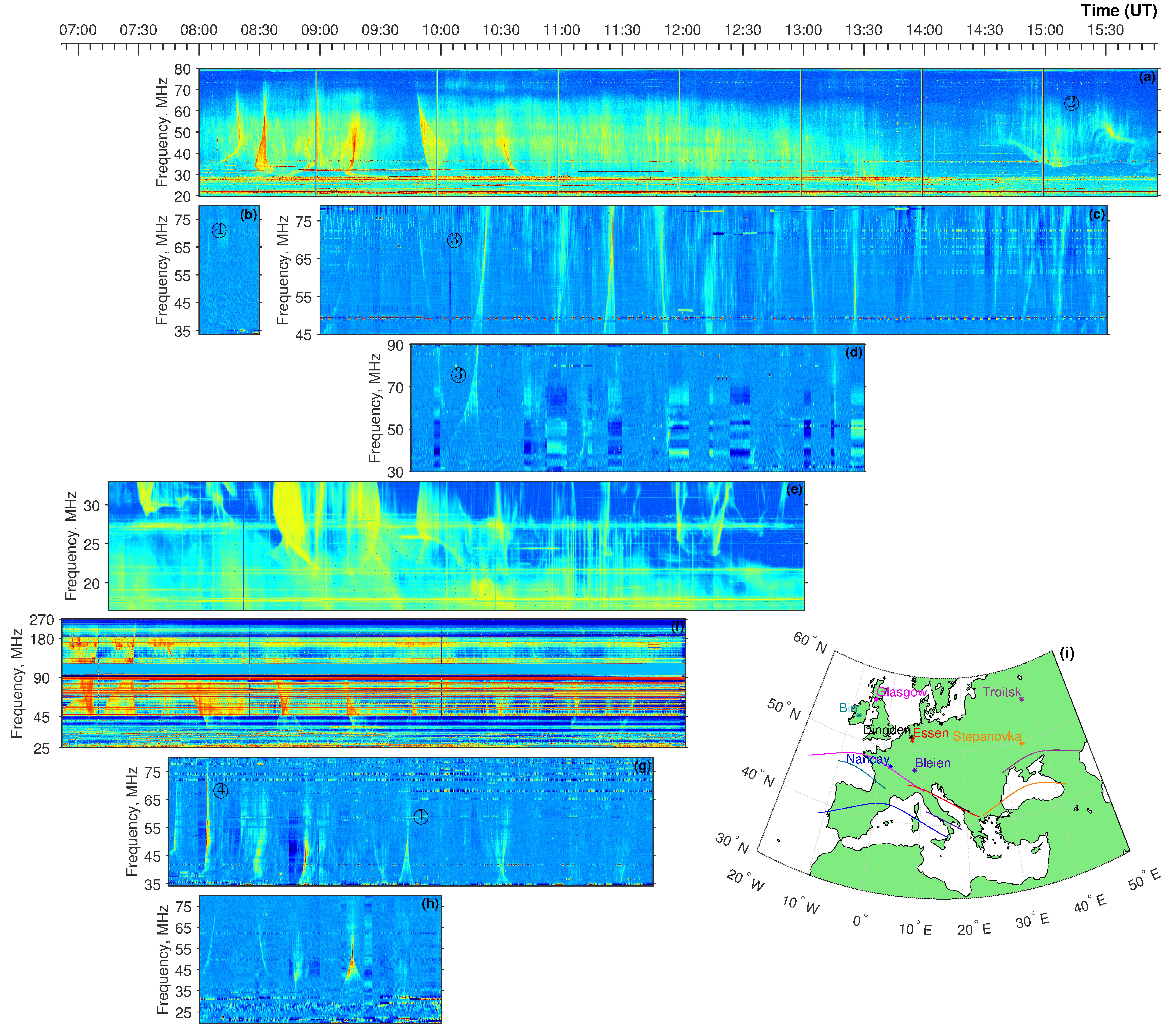}
\caption{Composition of the time-aligned solar dynamic spectra from NDA (20-80 MHz) (a), CALLISTO DARO (34-80 MHz) (b), CALLISTO GLASGOW (45-80.8 MHz) (c), CALLISTO BIR (30-90.3 MHz) (d), URAN-2 (16.5-33.0 MHz) (e), IZMIRAN (25-270 MHz) (f), CALLISTO ESSEN (34-80 MHz) (g), CALLISTO BLENSW (19.7-79.6 MHz) (h) on January 8, 2014. The common time axis is in the range 06:52~UT - 15:55~UT. The scales of frequency axes in dynamic spectra are arbitrary. The numbers -- from 1 to 4 -- inscribed in circles label four cases with SCs that are examined in the present study. (i) The map of the antennas locations and IPP tracks associated with the solar observational sessions. The observing sites and the corresponding IPP paths are marked by the same color.}
\label{Figure2}
\end{center}
\end{figure*}

Figure~\ref{Figure2}(a-h) shows a collection of the eight solar dynamic spectra on 2014 January 8. The antennas are scattered in Europe. The map of their locations is displayed in panel (i) of the figure. All instruments operated in overlapping frequency bands. In the spectrograms SCs can be recognized by their distinctive spectral shapes. Most SCs belong to the prevalent inverted V-like type. Some SCs appear in form of lanes, corresponding to the fiber-like type. A few SCs cannot be classified with the classification scheme defined by \citet{Koval2017}.

In Figure~\ref{Figure2}(i) we plotted the IPP trajectories for every Sun-antenna pair using equation~\ref{Eq1}. The coordinates of antennas are known. The solar zenith and azimuth angles were extracted with the high-precision solar position algorithm \citep{Reda2004}. The lengths of IPP tracks depend on the time spans of the corresponding dynamic spectra shown in
Figure~\ref{Figure2}(a-h). There are nearly 50 SCs in Figure~\ref{Figure2}. Below we will consider several cases to demonstrate the method of our study.


\subsection{Case $\#$1}


\begin{figure*}[ht]
\begin{center}
\includegraphics[width=0.8\textwidth]{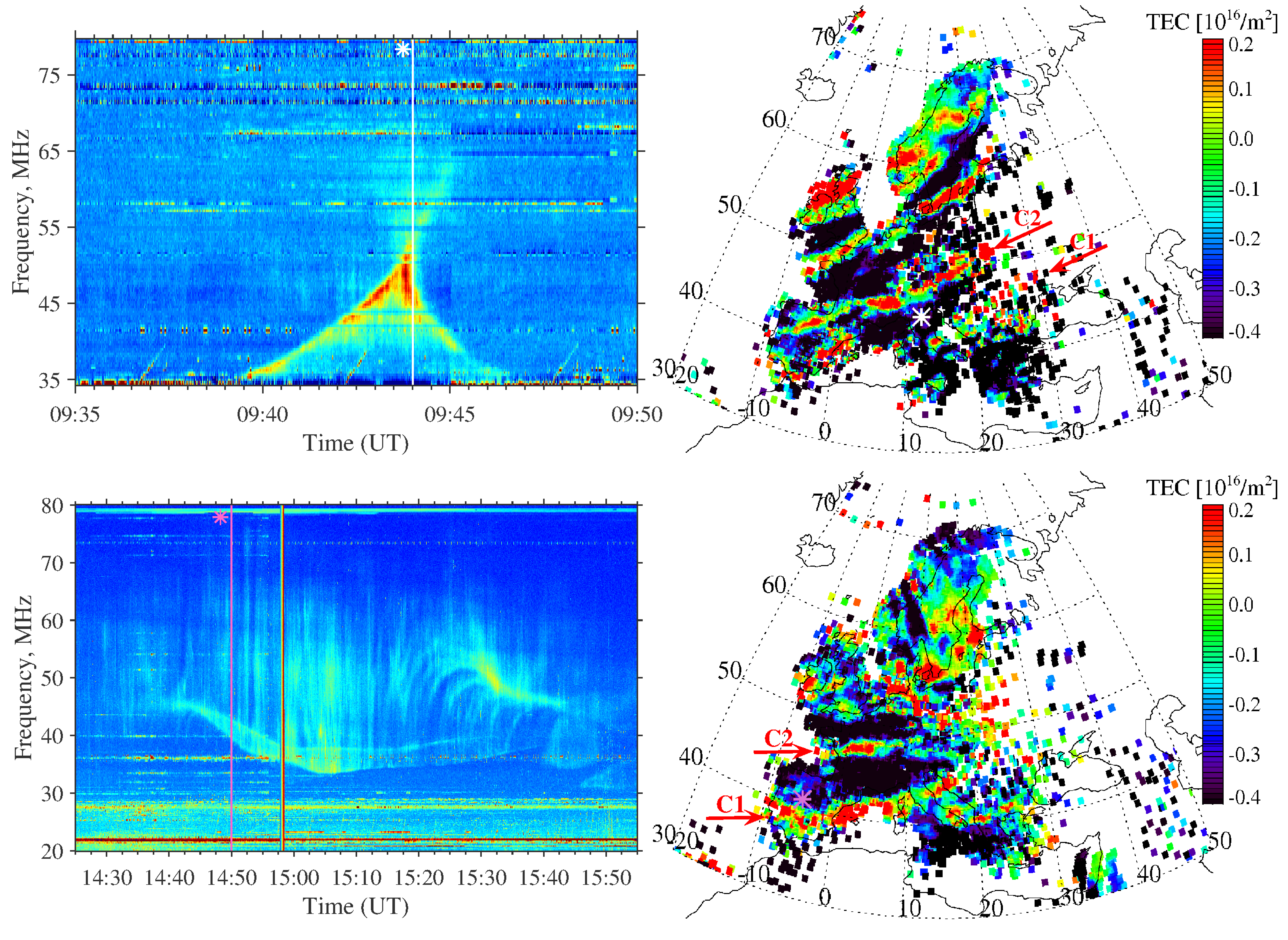}
\caption{(\textit{Top}) The CALLISTO ESSEN dynamic spectrum with the SC recorded on January 8, 2014. The white solid line on the spectrogram is at 09:44~UT. On the dTEC map on the right, the white asterisk shows the location of the IPP that is in the middle of the valley enclosed by two crests C1 and C2 (red arrows). The map and the IPP coordinates are obtained for the same time instant. (\textit{Bottom}) The NDA dynamic spectrum including the SC detected on January 8, 2014. The purple solid line on the spectrogram is at 14:50~UT. On the dTEC map on the right, the purple asterisk points the position of the IPP, which is within the valley between two crests C1 and C2 (red arrows). The map and the IPP coordinates are obtained for the same time instant.\smallskip\\
An animation of this figure is available. It runs from 09:39:00~UT to 09:46:30~UT in the top portion and from 14:40:00~UT to 15:50:00~UT in the bottom part of the animation. The video duration is 7 seconds.}
\label{Figure3}
\end{center}
\end{figure*}

In the upper part of Figure~\ref{Figure3} the SC recorded by the CALLISTO ESSEN is presented. This SC is almost symmetrical and likely belongs to the X-like type. It consists of two bright envelopes, crossing in frequency near 50 MHz, and its upper part is diffusive. The position of the IPP at 09:44:00~UT (near the SC's symmetry line) was determined and plotted on the dTEC map. It is seen that the IPP (white asterisk) locates in the center of the valley between two crests C1 and C2 as pointed by red arrows.

The upper part of Figure~\ref{Figure3} represents a still frame of the top part of the animation, which shows the obtained IPP positions imposed on dTEC maps from 09:39:00~UT to 09:46:30~UT with a 30-sec cadence. This time range corresponds to the longest SC duration at the lowest frequency ($\sim35$ MHz) in the spectrogram. It can be seen that during this time all IPPs lie within the valley of the moving wave-like structure. The SC could still exist below 35 MHz with a longer duration. In this case, even assuming the duration of the SCs is twice longer than at 35 MHz, the thus-obtained IPPs would still remain being in the valley. This SC can be regarded as a classical manifestation of the focusing of solar radiation caused by MSTIDs.

\subsection{Case $\#$2}

Unlike the previous case, the one in the bottom part of Figure~\ref{Figure3} is characterized by non-classified spectral structures. Besides, the total duration of the event is more than 1 h, that makes it rare too. Seemingly, there are two long-living SCs overlapping in time. The first SC starts at 14:40~UT in time and at 45 MHz in frequency. It takes a tunnel-like form, which firstly goes down in frequency to 35 MHz and then keeps a slight slope from 15:00~UT up to the end at 15:40~UT. The second SC has a curved backbone-like structure with a number of filaments extended out from it. Tentatively, this SC may be attributed to a combination of the fiber-like and the fringe-like types. The SC's filaments firstly appear near 15:18~UT. From 15:40~UT the SC becomes diffusive.

The position of the IPP taken at 14:50~UT was put on dTEC map which is on the right side at the bottom of Figure~\ref{Figure3} (see purple asterisk). It is clearly seen that the IPP locates near the crest C1. The wavelength of this TID -- the distance between the crests C1 and C2 (red arrows) -- is approximately 550 km. In the bottom portion of the Figure 3 animation, one can observe the IPP trajectory on dTEC maps from 14:40~UT to 15:50~UT. At 14:40~UT it starts from the position very close to the crest C1, and later reaches the crest C2 at 15:50~UT. On a closer examination, the IPP track is basically along the TID wave front. So the track remains at the valley for a long time. This may explain the long durations of the SCs as well as their uncommon spectral shapes. Evidently, both SCs in the considered case were produced by the same TEC perturbation.

\subsection{Case $\#$3}


\begin{figure*}[ht]
\begin{center}
\includegraphics[width=0.8\textwidth]{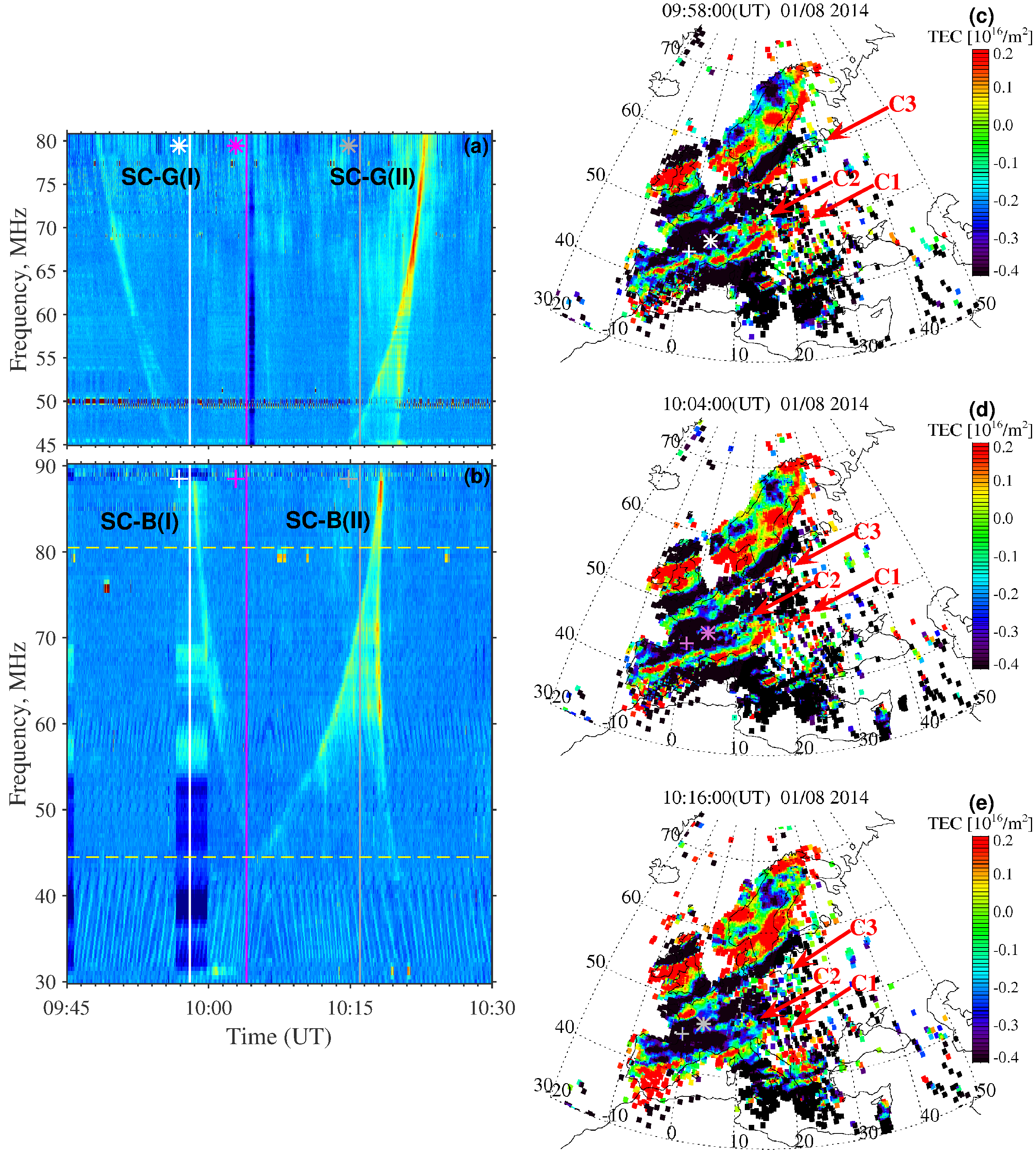}
\caption{The CALLISTO GLASGOW (a) and the CALLISTO BIR (b) dynamic spectra with two pairs of SCs, SC-G(I,II) and SC-B(I,II), respectively, observed on January 8, 2014. The frequency axes of the dynamic spectra are in the same scale. The dashed horizontal lines on the spectrogram in panel (b) show the frequency range of the spectrogram in panel (a). The vertical white, purple, and gray lines indicate three time instants at 09:58~UT, at 10:04~UT, and at 10:16~UT, correspondingly, in both panels (a) and (b). The IPPs relating to the CALLISTO GLASGOW (asterisk) and the CALLISTO BIR (plus) spectrograms for these instants were obtained and plotted on dTEC maps in panels (c-e). The corresponding lines and marks have the same colors. The crests C1, C2, and C3 are pointed by red arrows on the dTEC maps.\smallskip\\
An animation is available. It begins on 09:48:00~UT and ends on 10:24:00~UT. The duration of the video is 7 seconds.}
\label{Figure4}
\end{center}
\end{figure*}

Figure~\ref{Figure4} presents the SCs which are likely caused by the same TID and recorded by different radio telescopes with similar spectral morphology. Panels (a) and (b) of the figure represent the CALLISTO GLASGOW and the CALLISTO BIR solar spectrograms, respectively. The dashed horizontal lines on the CALLISTO BIR spectrogram mark the band of the CALLISTO GLASGOW spectrogram in frequency. Three time instants at 09:58~UT (white line), 10:04~UT (purple line), and 10:16~UT (gray line) in both dynamic spectra have been indicated. The IPPs corresponding to the CALLISTO GLASGOW (asterisk) and the CALLISTO BIR (plus) spectrograms for these instants were obtained and plotted on dTEC maps. The corresponding lines and signs have the same colors. Two pairs of SCs in the dynamic spectra of the CALLISTO GLASGOW and BIR stations are denoted as SC-G (I,II) and SC-B (I,II), correspondingly. The animation of this case is available with Figure~4.

The SC-G(I) in panel (a) starts near 09:48~UT at 80.8 MHz. It is a lane drifting from high to low frequencies. During the SC-G(I) life time (09:48~UT-09:58~UT) the successive IPPs lie within the valley between the crests C1 and C2 (red arrows). The IPP taken at 09:58~UT approaches the crest C2 (white asterisk in panel (c)). Because of limited frequency range, some parts of the SC-G(I) are not observable. Nevertheless, one can compare it with the SC-B(I) in panel (b), which begins in time with some delay. Obviously, both SCs look almost identical. However, time lag between their onset times at 80.8 MHz is about 11 minutes. Let us examine the panels (c), (d) and (e). Due to favorable observing conditions, the pairs of IPPs at any given moment are in alignment with the TID wavefronts. This explains the possibility of almost simultaneous detection of similar SCs in different places. At the same time, the crest C1 looks not straight along its length. In panel (c) it has a noticeable break point near $5^\circ$ in longitude that divides the crest on eastern and western parts. Thus, the eastern part shows a shear in the moving direction relative to the western part. Both parts propagate synchronously though (see the animation in Figure 4). This spatial displacement likely results in the time lag between the SC-G(I) and the SC-B(I). Indeed, from 09:48~UT to 09:58~UT -- the SC-G(I) life time -- the IPPs corresponding to the CALLISTO BIR antenna locate exactly on the western part of the crest C1. Only at 09:58~UT, when the IPP reaches the valley (white plus on dTEC map in panel (c)), the SC-B(I) appears.

Figure~\ref{Figure4}(d) demonstrates the IPP positions at 10:04~UT. The IPP associated with the CALLISTO BIR antenna is in the valley (purple plus), while the IPP associated with the CALLISTO GLASGOW antenna is on the crest C2 (purple asterisk). On the dTEC map the crest C2 is partly depressed within $-6^\circ$--$5^\circ$ range of longitudes. Its short fragment is observable from $-8^\circ$ to $-6^\circ$ in longitude. Thus, two valleys in between the crests C1, C2 and C3 can be seen along latitude and in the longitude belt from $5^\circ$ to $11^\circ$. However, there is only one comparably wider valley between the crests C1 and C3 within $-6^\circ$--$5^\circ$ range of longitudes. The IPP track of the CALLISTO GLASGOW antenna crosses two valleys. As a result, the SC-G(I) and SC-G(II) occur with difference in time. The IPP track of the CALLISTO BIR antenna moves across one valley. Interestingly, two SCs -- the SC-B(I) and SC-B(II) -- emerge as well but overlap in time. We have already described similar occurrence in the previous case in the NDA dynamic spectrum. The possible interpretation will be given in section~\ref{Sect:Summary and Discussion}.

In Figure~\ref{Figure4}(e) the IPP locations at 10:16~UT are shown. The IPP belonging to the CALLISTO BIR antenna locates in the wide valley (grey plus) between the crests C1 and C3. The IPP pertaining to the CALLISTO GLASGOW antenna is in the valley (grey asterisk) between the crests C2 and C3, closer to the former. In panels (a) and (b), this time corresponds to the moment right after the onset of the SC-G(II), while the SC-B(II) has lasted almost half of its lifetime by then. The SC-B(II) ends at 10:21~UT. The SC-G(II) disappears later near 10:24~UT. The respective IPPs, corresponding to these time moments, have reached the crest C3, and as a result, the focusing action of this TEC perturbation was completed (see the animation in Figure 4). Visually, the SC-G(II) and the SC-B(II) differ more than the previous pair. Nevertheless, the performed analysis of the dTEC maps confidently confirms their common origin. The computed spacing between two tracks of IPPs at the 250-km height is about 370 km. The distinctions in spectral shapes of the SC-G(II) and the SC-B(II) might have arisen from variations of the focusing conditions such as magnitude of electron density, height and width of TID layer, spatial structure of valley and crests in different locations of the same TEC perturbation.

\subsection{Case $\#$4}


\begin{figure*}[ht]
\begin{center}
\includegraphics[width=0.8\textwidth]{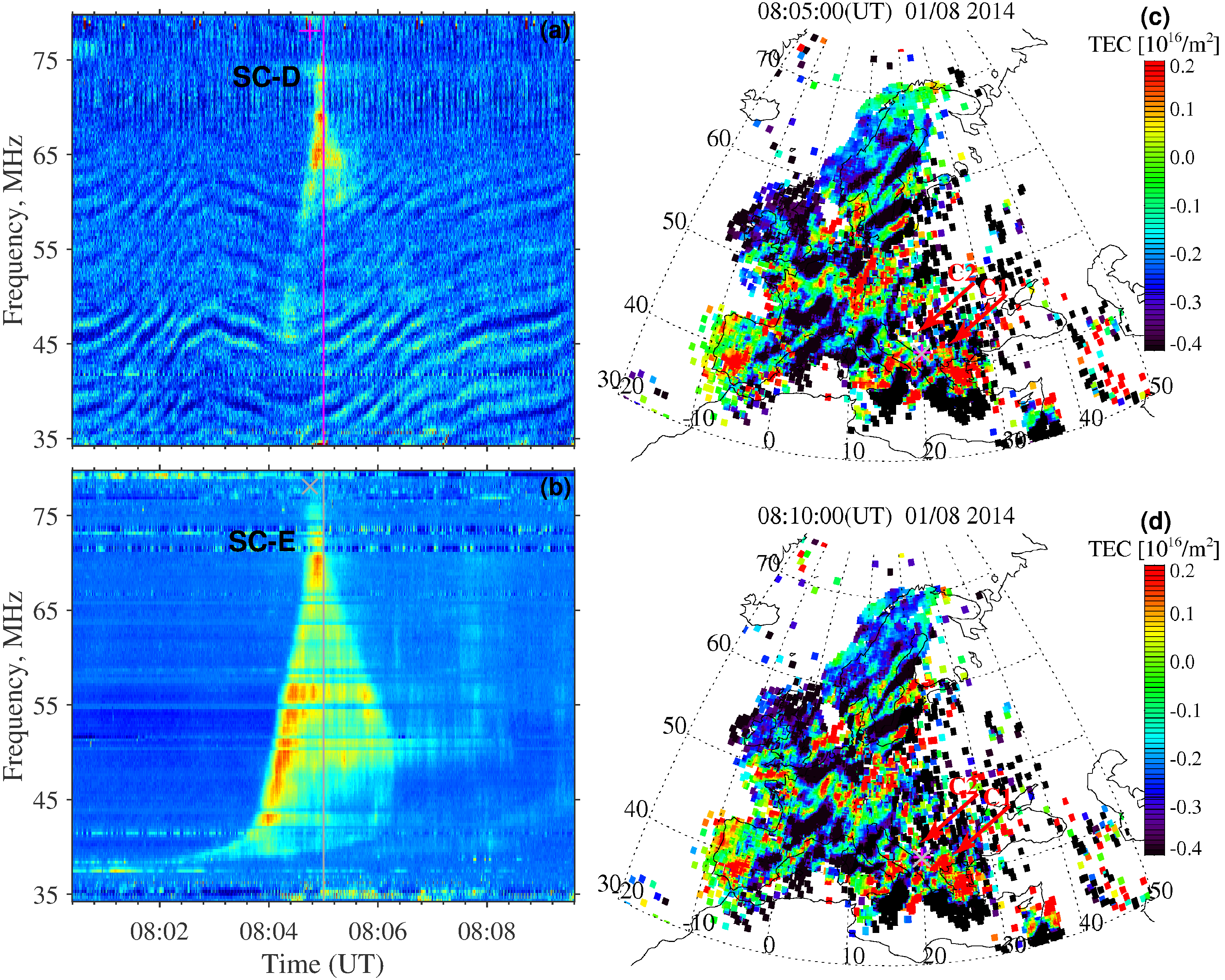}
\caption{The CALLISTO DARO (a) and the CALLISTO ESSEN (b) dynamic spectra with SCs, SC-D and SC-E, respectively, recorded on January 8, 2014. The purple and grey lines in both spectrograms indicate the time moment at 08:05~UT. The IPPs for this instant are marked by purple plus (the SC-D) and grey cross (the SC-E) on the dTEC map in panel (c). In panel (d), the same IPPs positions were imposed on the dTEC map obtained at 08:10~UT. The crests C1 and C2 are shown by red arrows in both dTEC maps.}
\label{Figure5}
\end{center}
\end{figure*}

In Figure~\ref{Figure5} we show a pair of SCs which have been detected by two spatially spaced instruments. This represents a rare situation. In contrast to the previous case, here the SCs are almost replicas of each other. Panels (a) and (b) of the figure represent the CALLISTO DARO and the CALLISTO ESSEN solar dynamic spectra, respectively. While the SC in the CALLISTO ESSEN spectrogram (SC-E) is enhanced, the SC in the CALLISTO DARO spectrogram (SC-D) is partly suppressed in intensity. The upper part, front and back envelopes of the SC-D stand out distinctly though. Both SCs belong to the inverted V-like type. Moreover, they coincide in appearance time as well as frequency range.

We selected the instant at 08:05~UT near the SCs' symmetry lines. This moment is indicated by purple and grey lines in panels (a) and (b), respectively. The IPPs at 08:05~UT for the CALLISTO DARO (purple plus) and the CALLISTO ESSEN (grey cross) antennas were computed and plotted on the dTEC map in panel (c). The IPPs locate in the valley bordering by the crests C1 and C2 (red arrows). The IPP marks overlap with each other. It should be noted that the distance between the observing sites is around 48 km. It is the shortest distance between radio observatories from which data are used in this study. The calculated distance between the IPPs at the 250-km altitude is nearly 17 km during the event. It means that the focusing conditions are almost the same, resulted in almost identical SCs.

In Figure~\ref{Figure5}(c) the crests C1, C2 and the valley are hardly recognized. We fix the coordinates of IPPs for the instant at 08:05~UT. Then, we put them on dTEC map obtained at 08:10~UT in panel (d). There the same crests C1, C2 and the valley can be discerned clearly. We can easily demonstrate that both crest-valley-crest structures in panels (c) and (d) are the same. If we assume that TID velocity is of 200 m/s, then in 5 mins the TID wavefront would propagate at a distance of about 60 km. This distance is within one pixel of the dTEC map. Thus, we relate the TEC perturbation to the SCs.

\subsection{Uncertainties in IPP positions}

There are two obvious uncertainty sources in the determination of IPP coordinates on dTEC map. The first one is the solar position. To get the solar zenith and azimuth angles at each observing site for equation~\ref{Eq1}, we used online routine (\url{https://midcdmz.nrel.gov/solpos/spa.html}) described in \citep{Reda2004}. It calculates solar coordinates in the period from the year -2000 to 6000. The declared uncertainties are of $\pm0.0003^\circ$. Therefore, the position error of an IPP arising from this routine is negligible.

The second source of uncertainties in an IPP location on dTEC map is related to the effect of ionospheric refraction on radio ray propagation. The expressions for latitude and longitude of an IPP in equation~\ref{Eq1} are frequency-independent. Indeed, they are valid in the GPS measurements where electromagnetic waves at gigahertz range are exploited. The ionosphere is almost transparent in this frequency range. In our study, radio observations below 300 MHz were analyzed and the ionosphere affects the wave propagation. We can estimate the angular deviation of a radio ray introduced by the ionosphere with TIDs. \citet{Bougeret1981} collected the peak-to-peak angular positional fluctuations from observations of different sources: solar sources (type I bursts and quiet Sun), strong radio sources (Cygnus A, Cassiopea), and radio beacons of space satellites. The fluctuations were believed to be caused by TIDs. The author derived its maximum amplitude $\theta_m\simeq5\cdot10^4f^{-2}$, where $\theta_m$ in arc min, $f$ in MHz. It is seen that $\theta_m$ decreases with increasing frequency. The SCs examined above are all observed above 35 MHz. This frequency can be selected to obtain the upper limit of deviations introduced by TIDs. At 35 MHz, the deviation is about 41 arc mins that is comparable to the pixel size ($45^\prime\times45^\prime$) of dTEC map. Therefore, we conclude that the IPP coordinates can only be slightly affected with deviations smaller than the spatial resolution of TIDs in the considered cases.

\section{Summary and Discussion}
\label{Sect:Summary and Discussion}

In this study, for the first time we provided strong observational evidence of cause-effect relationship between pairs of TIDs and SCs. In all examined cases, the performed timing analysis allowed us to reveal the couplings between MSTIDs and SCs with great confidence. We showed that the observed peculiarities of the spatial structure of TIDs on dTEC maps were consistent with the time-frequency characteristics of the corresponding SCs in the spectrograms. In addition, the study settled a controversy about the origin of SCs in the solar radio records, which may be mistakenly attributed to the Sun. If so, the events would present exact morphologies and occurrence times regardless of the location of radio observatories. In this work, we have demonstrated a collection of meter-decameter solar dynamic spectra obtained by various radio instruments around Europe. It is seen that SCs occur almost uniquely in each dynamic spectrum. Thus, the peculiar features of SCs are not of solar origin, but resulted from the varying ionosphere irregularities through which the solar radiation propagates.

The above-mentioned statement is further supported by the observations of SCs in cases $\#$3 and $\#$4. In case $\#$3, two pairs of very similar SCs were recorded. We found that the focusing was initiated by the same TEC perturbation. However, existing variations of parameters within it brought distinctions in the respective SCs. In case $\#$4, we established that because of the vicinity of radio observatories, both antennas received the solar radiation which passed through the same region of the TID valley. So, identical SCs were detected.

It is necessary to clarify the appearances of overlapping SCs, which are caused by single TID valleys like in cases $\#$2 and $\#$3. It is expected that one TID valley may cause one caustic surface in space which is registered as a SC at one observing place. We suppose that this scenario is correct for pure TID valleys. Presumably, if a TID valley itself contains fluctuations of electron density, more than one caustic surface may be generated. On dTEC maps such fine structure may be poorly distinguishable or considered as not-significant. In TID structures in cases $\#$2 and $\#$3, the valleys do comprise some minor and isolated irregularities with the crests bordering the valleys. Likely, these irregularities account for double occurrences of spatial caustics resulted in partial overlap in time of spectral caustics.

We should also emphasize the importance of the paper for solar radio observations in light of realization of new ambitious projects of grand building of radio astronomy facilities. Thus, in the recent years the global trend in the deployment of radio astronomy instruments tends towards broadband antenna arrays (NenuFAR, LOFAR, GURT, MWA, LWA, SKA), operating in meter-decameter wavelength range. In this connection, the subject of transients of various origins in HF and VHF solar observations is regarded as of current relevance.

This research is motivated by our previous works \citep{Koval2017,Koval2018}. With this series of studies on SCs, we pursue a better understanding of this intriguing phenomenon. Further investigations will be continued.

\acknowledgments
This research was supported by NNSFC grants 11790303 (11790300), 41774180, and 11750110424. The authors are grateful to the Nan\c{c}ay team for their open access data policy. Also, the authors thank the teams who support and operate various CALLISTO stations. GPS data were provided by IGS, UNAVCO, SOPAC, EPN, BKGE, OLG, IGNE, DUT, ASI, ITACYL, ESEAS, SWEPOS, NMA, BIGF, RENAG, and ITINGV. GPS-TEC database was partially supported by JSPS KAKENHI Grant Number JP16H06310, JP15H05813, and JP16H06286.

\end{document}